\documentclass[prl,twocolumn,superscriptaddress,showpacs,amsmath,amssymb]{revtex4}
\usepackage{graphicx}

\newcommand{\ket}[1]{\, | #1 \rangle}

\newcommand{\hb}{\hat{b}}
\newcommand{\hc}{\hat{c}}
\newcommand{\hbd}{\hat{b}^{\dagger}}
\newcommand{\hcd}{\hat{c}^{\dagger}}
\newcommand{\hn}{\hat{n}}

\newcommand{\be}{\begin{equation}}
\newcommand{\ee}{\end{equation}}
\newcommand{\bea}{\begin{eqnarray}}
\newcommand{\eea}{\end{eqnarray}}
\newcommand{\bean}{\begin{eqnarray*}}
\newcommand{\eean}{\end{eqnarray*}}
\newcommand{\besa}{\begin{subequations}\begin{eqnarray}}
\newcommand{\eesa}{\end{eqnarray}\end{subequations}}
\newcommand{\sgn}{\operatorname{sgn}}
\renewcommand{\mod}{\operatorname{mod}}

\begin{document}
\title{Three-Body Bound States in a Lattice}
\author{Manuel Valiente}
\affiliation{Institute of Electronic Structure and Laser, FORTH, 
71110 Heraklion, Crete, Greece}
\affiliation{Institut f\"ur Physik, 
Humboldt-Universit\"at zu Berlin, Hausvogteiplatz 5-7, D-10117 Berlin, Germany}
\author{David Petrosyan}
\affiliation{Institute of Electronic Structure and Laser, FORTH, 
71110 Heraklion, Crete, Greece}
\author{Alejandro Saenz}
\affiliation{Institut f\"ur Physik, 
Humboldt-Universit\"at zu Berlin, Hausvogteiplatz 5-7, D-10117 Berlin, Germany}
\date{\today}

\begin{abstract}
We pursue three-body bound states in a one-dimensional tight-binding 
lattice described by the Bose-Hubbard model with strong on-site interaction.
Apart from the simple strongly-bound ``trimer'' state corresponding to all 
three particles occupying the same lattice site, we find two novel kinds 
of weakly-bound trimers with energies below and above the continuum of 
scattering states of a single particle (``monomer'') and a bound particle 
pair (``dimer''). The corresponding binding mechanism can be inferred 
from an effective Hamiltonian in the strong-coupling regime which contains
an exchange interaction between the monomer and dimer. In the limit of 
very strong on-site interaction, the exchange-bound trimers attain 
a universal value of the binding energy. These phenomena can be 
observed with cold atoms in optical lattices.
\end{abstract}


\maketitle


\paragraph{Introduction.}

For decades, many-body physics in periodic potentials has been one of 
the central topics of research in condensed-matter theory \cite{SolStPh}.
The complexity of such systems requires the use of model
Hamiltonians which, for sufficiently deep lattices, take into account
only short-range interactions and neglect coupling between different 
bands of the periodic potential \cite{Hubbard}. Despite apparent
simplicity, tight-binding many-body Hamiltonians, such as the Hubbard 
model, have been able to elucidate and predict many important lattice 
phenomena. A prime example is the transition from a superfluid to 
a Mott insulator phase \cite{bosMI} experimentally realized with 
cold bosonic atoms in optical lattices \cite{optlattMI}. In another 
recent experiment \cite{KWEtALPZ}, stable repulsively-bound pairs of atoms
in an optical lattice were observed. This seminal achievement has led to 
several theoretical studies of the dynamics of atom pairs in periodic 
potentials \cite{molmer,MVDPdmrs,MVDPexH,fshbchRes,DPKLT,weiss,WHCh,JChS}. 

Two decades ago, Mattis \cite{Mattis} established an equation 
for the bound states of three bosons, and predicted 
Efimov states \cite{Efimov} in a 3D lattice. 
Here we study the three-body problem in a one-dimensional (1D) 
tight-binding lattice described by the Bose-Hubbard model with 
nearest-neighbour hopping and strong on-site interaction. 
We find that, in general, for each value of total quasi-momentum 
there are three distinct three-body bound states. 
The first strongly-bound ``trimer'' state is simple, 
corresponding to all three particles being at the same lattice site. 
The other two hitherto unknown bound states arise due to an 
effective exchange interaction between a bound 
pair---``dimer''---and a single particle---``monomer''. 
These weakly-bound trimers have energies slightly below and above
the two-body continuum of the monomer and dimer scattering states. 
All three trimer states exist irrespectively of the sign of  
on-site interaction, while for very strong interaction, the 
exchange-bound trimers attain a universal value of the binding 
energy equal to half the single particle hopping rate.


\paragraph{Mathematical formalism.}

We consider bosonic particles in a 1D lattice described 
by the Hubbard Hamiltonian     
\be
H = - J \sum_{j} (\hbd_j \hb_{j+1} + \hbd_{j+1} \hb_j ) 
+ \frac{U}{2} \sum_{j} \hat{n}_j(\hat{n}_j-1) , \label{BHH}
\ee
where $\hbd_{j}$ ($\hb_{j}$) is the particles creation (annihilation) 
operator and $\hn_j = \hbd_{j} \hb_{j}$ the number operator at $j$th 
lattice site, $J$($>0$) is the hopping rate, and $U$ is the on-site
interaction (which can be attractive or repulsive). 
We seek three particle bound states in momentum representation,
\be
\ket{\psi} = \frac{1}{(2\pi)^{3/2}} \iiint_{\Omega^3} \!\! dk_1 dk_2 dk_3 \,
\psi(k_1,k_2,k_3) \ket{k_1,k_2,k_3},
\ee
where the wavefunction $\psi(k_1,k_2,k_3)$ is symmetric with respect to 
exchange of any pair of particles, with each particle quasi-momentum 
$k_j \in \Omega$ restricted to the first Brillouin zone 
$\Omega \equiv  [-\pi,\pi]$. From the stationary Schr\"odinger equation 
$H\ket{\psi}=E\ket{\psi}$, using the conservation of total quasi-momentum
$K=k_1+k_2+k_3 \; (\mod 2\pi)$, we obtain \cite{Mattis}  
\be
\psi(k_1,k_2,k_3) = -\frac{M(k_1)+M(k_2)+M(k_3)}
{\epsilon(k_1)+\epsilon(k_2)+\epsilon(k_3)-E},
\ee
where $\epsilon(k)=- 2J\cos(k)$ is the single-particle energy (Bloch) band 
of Hamiltonian (\ref{BHH}), while the functions $M(k)$ satisfy the 1D
Mattis integral equation \cite{Mattis}
\bea
&& M(k) [1+I_E(k)] \nonumber \\ &&
= -\frac{U}{\pi}\int_{-\pi}^{\pi} \!\! dq \, 
\frac{M(q)}{\epsilon(k) + \epsilon(q) + \epsilon(K-k-q) - E }, \label{Minteq}
\eea
with $I_E(k)$ being a generalized Watson integral \cite{Watson}
\bean
I_E(k) &\equiv& \frac{U}{2\pi} \int_{-\pi}^{\pi}
\!\! dq \, \frac{1}{\epsilon(k) + \epsilon(q) + \epsilon(K-k-q)-E} \\
&=&-\frac{\sgn [ E -\epsilon(k) ] U}
{\sqrt{ [ E-\epsilon(k) ]^2-16J^2\cos^2[(K-k)/2]}}.
\eean
Equation (\ref{Minteq}) can be cast as a homogeneous Fredholm 
equation of the second kind with eigenvalue $\lambda=1$. 
Hence, for a given $U/J$ and fixed $K$, it is a nonlinear 
equation for energy $E$, which can be solved numerically. 
Note also that Eq.~(\ref{Minteq}) does not imply $M(k)$ to be 
an even or odd function; in fact, there is no symmetry 
with respect to $k=0$, unless $|K|= 0$ or $\pi$.


\paragraph{Three-body spectrum.}

As is well-known \cite{Mattis}, the Bose-Hubbard Hamiltonian (\ref{BHH})
with $|U|/J \gg 1$ has a narrow band of {\it on-site} bound states, 
corresponding to $N>1$ particles tightly co-localized on the same 
lattice site, with the energy $E_b \approx \frac{1}{2}U(N^2-N)$. 
In the present case of $N=3$, we have $E_b \approx 3U$. 
The other two known parts of the spectrum of Eq.~(\ref{BHH}) are: 
the three-body scattering continuum of three (asymptotically) 
free particles, with the energy given by the sum of single-particle bands, 
$E_{c3} = \epsilon(k_1) + \epsilon(k_2)+ \epsilon(K-k_1-k_2)$; 
and the two-body scattering continuum of a bound pair 
(dimer) and a free particle (monomer), with energy 
$E_{c2} =\sgn(U)\sqrt{U^2+ [4 J \cos(Q/2)]^2} - 2J \cos(K-Q)$, where
the first term is the energy of a dimer with quasi-momentum $Q$ 
\cite{KWEtALPZ,molmer,MVDPdmrs}.

\begin{figure}[t]
\includegraphics[width=0.48\textwidth]{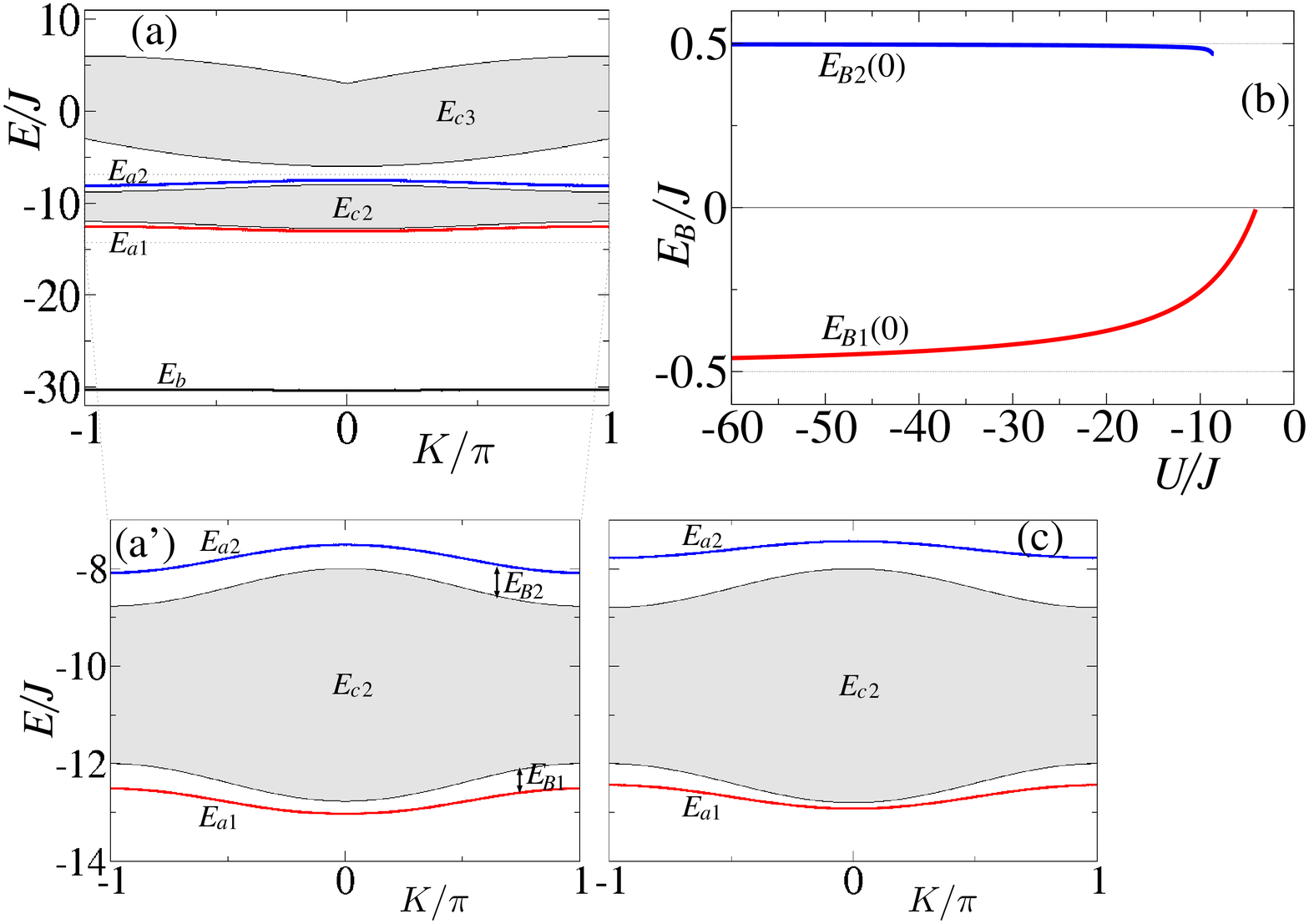}
\caption{(a) Full three-particle energy spectrum 
of Hamiltonian (\ref{BHH}) with $U=-10J$, 
versus the total quasi-momentum $K$.
All bound states are obtained via 
exact numerical solution of Eq.~(\ref{Minteq}). 
(a') Magnified part of the spectrum corresponding to dimer-monomer states. 
(b) Binding energies $E_B$ for the off-site (weakly-bound) 
trimers at $K=0$ versus the interaction strength $U <0$.
(c) Dimer-monomer spectrum of the effective Hamiltonian (\ref{effHam}).
The two bound states are obtained via numerical solution of Eq.~(\ref{dmPsi}).}
\label{fig:spectrum}
\end{figure}

The full spectrum of Hamiltonian (\ref{BHH}) 
is shown in Fig.~\ref{fig:spectrum}(a). 
For concreteness, here we consider attractive interaction, $U<0$, 
but note that our results equally apply to the case of repulsive 
interaction, $U>0$ \cite{cmntU-U}. 
As seen in the figure, lattice bosons can form two new kinds of 
three-body bound states whose energies $E_{a1}$ and $E_{a2}$ lie 
below and above the two-body continuum $E_{c2}$.
Some properties of these states can be deduced by energy considerations. 
First, these are not on-site bound states, 
since their energies $E_{a1(2)} \simeq U + O(J)$ are far from $3U$. 
Next, their binding energies, with respect to the  $E_{c2}$ band, 
are $E_{B1(2)} \sim \mp J/2$ [see Fig.~\ref{fig:spectrum}(a'),(b)], 
which suggests that these are {\it off-site} weakly-bound states 
of a dimer and a monomer. Note that the state above the two-body 
continuum is bound stronger than the state below the continuum. 
Finally, they are not Efimov states which can exist only in 3D 
systems near two-boson resonance \cite{Mattis,Efimov,Nielsen},
since we examine a 1D lattice with no two-particle scattering 
resonances \cite{2bres}. 

\begin{figure}[t]
\includegraphics[width=0.48\textwidth]{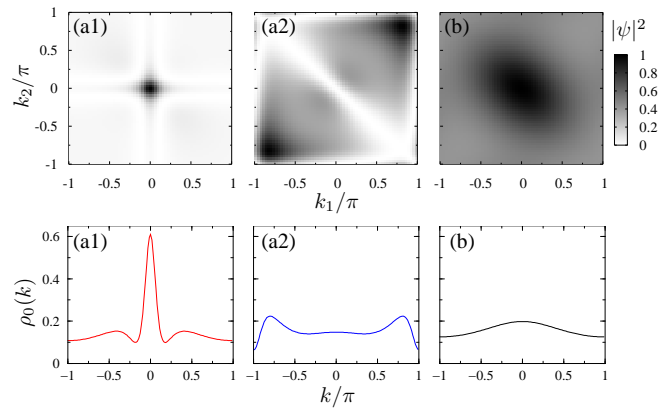}
\caption{Top panel: Quasi-momentum distributions $|\psi(k_1,k_2,K-k_1-k_2)|^2$
of the three-particle bound states for $U=-10J$ and $K=0$; 
(a1) and (a2) correspond to the off-site (weakly-bound) trimers 
with energies below and above the two body dimer-monomer continuum, 
while (b) corresponds to the on-site (strongly-bound) trimer.
Lower panel: Reduced single-particle momentum distributions $\rho_{0}(k)$
for the cases of (a1), (a2) and (b).}
\label{fig:wavefunct}
\end{figure}

In Fig.~\ref{fig:wavefunct} we show the quasi-momentum distributions 
$|\psi(k_1,k_2,K-k_1-k_2)|^2$ for the three bound states at total 
quasi-momentum $K=0$, as well as the corresponding reduced 
single-particle momentum distributions 
$\rho_{K}(k) = \int_{\Omega} \! dq \, |\psi(k,q,K-k-q)|^2$. 
In cold atom experiments \cite{OptLatRev}, $\rho_{K}(k)$ can 
be directly mapped out by absorption imaging of the atomic density 
distribution after sudden release from the lattice potential 
followed by a time-of-flight. For the trimer state with energy 
$E_{a1}$ lying below the two-body continuum $E_{c2}$, the Mattis 
function $M(k)$ at $K=0$ is symmetric, and the corresponding 
quasi-momentum distribution is peaked at $k=0$, similar to that 
of an attractively bound pair of particles \cite{molmer,MVDPdmrs}. 
For the bound state with energy $E_{a2}$ above the continuum, 
the $K=0$ Mattis function is antisymmetric with respect to $k=0$; 
the quasi-momentum distribution is peaked at $|k| \lesssim \pi$
resembling that of a repulsively bound particle pair 
\cite{KWEtALPZ,molmer,MVDPdmrs}. 
Finally, the on-site bound state (ground state for $U<0$) has quite
a flat momentum distribution, since the very large binding energy 
$E_b - E_{c2} \sim 2U$ makes the on-site trimer practically immobile. 

As can be seen from Fig. \ref{fig:spectrum}(b), where we plot 
the binding energies $E_{B1(2)}$ at $K=0$, there are thresholds
for the existence of full bands of the off-site bound states. 
For the trimer below the two-body continuum, the binding energy 
vanishes when $|U|\approx 4 J$: at this critical value of $U$
the trimer energy $E_{a1}$ approaches the edge of the dimer-monomer
scattering continuum $E_{c2} = - \sqrt{U^2 +16 J^2} -2 J$ 
resulting in $K=0$ scattering resonance. 
On the other hand, the trimer above the two-body continuum 
ceases to exist already for $|U| \approx 8.5 J$, since then 
its energy $E_{a2}$ approaches the bottom of the three-body 
continuum $E_{c3} = 3 \epsilon(0) = - 6J$ 
(the two continua, $E_{c2}$ and $E_{c3}$, overlap for $|U| \leq 8J$). 
Thus, at $K=0$, the trimer state with energy $E_{a2}$ starts to 
appear well in the strong interaction regime, while for larger $K$ 
the threshold is smaller: $|U| \approx 4 J$ for $|K| \to \pi$.


\paragraph{Effective model.}

The intriguing properties of the off-site trimers can be understood 
in term of an effective perturbative model valid in the strong
interaction regime ($|U|/J >8$). Upon adiabatically eliminating the 
on-site trimer states $\ket{3_j}$ and states $\ket{1_j1_k1_l}$ with 
all three particles at different lattice sites, for a single
dimer-monomer system we obtain an effective Hamiltonian 
to second order in $J$, which has the form 
\be
H_{\mathrm{eff}} = H_1 + H_2 + H_{\mathrm{int}} . \label{effHam} 
\ee
Here $H_1 = - J \sum_{j} (\hbd_j \hb_{j+1} + \mathrm{H.c.})$ describes 
the monomer with the single-particle spectrum $\epsilon(k)$. 
Next, 
\[
H_2 = \mathcal{E}^{(2)} \sum_{j} \hat{m}_j - 
J^{(2)} \sum_{j} (\hcd_j \hc_{j+1} + \mathrm{H.c.})
\]
is the Hamiltonian for a dimer, with $\hcd_j$ ($\hc_{j}$) being the dimer 
creation (annihilation) operator and $\hat{m}_j = \hcd_j \hc_j$ the number
operator at site $j$. The first term in $H_2$ is the dimer 
``internal energy'' $\mathcal{E}^{(2)} \equiv [U - 2 J^{(2)}]$, 
while the second term with the effective hopping rate 
$J^{(2)} = -2J^2/U$ gives rise to the dimer kinetic energy spanning
the Bloch band $\epsilon^{(2)}(Q) = - 2 J^{(2)} \cos(Q)$ \cite{MVDPdmrs,DPKLT}.
Finally, 
\[
H_{\mathrm{int}} =  V^{(2)} \sum_j \hat{m}_j\hat{n}_{j \pm 1}
- W \sum_j (\hcd_{j+1} \hc_j \hbd_j \hb_{j+1} + \mathrm{H.c.})
\]
describes effective short-range interactions between the 
dimer and monomer, including a weak repulsive (or attractive, if $U>0$)
nearest-neighbor interaction $V^{(2)} = -7 J^2/2U$, and an exchange
interaction with the rate $W = 2 J$, the latter being responsible
for the formation of off-site trimers as discussed below. 

In Fig.~\ref{fig:spectrum}(c) we plot the spectrum 
of the effective Hamiltonian~(\ref{effHam}). 
It involves a two-body scattering continuum, with the energy given by 
the sum of the energies of (asymptotically) free dimer and monomer, 
$E_{c2} = [U - 2 J^{(2)}] - 2 J^{(2)} \cos(Q) -2 J \cos (K-Q)$,
and two bound states with energies $E_{a1}$ and $E_{a2}$ 
below and above $E_{c2}$. These dimer-monomer bound states are 
obtained using the Schr\"odinger equation for the two-body wavefunction 
$\Psi(Q,k)$ in momentum space, which leads to the integral equation
\be
\Psi(Q,k)= -\frac{1}{2\pi} \int_{-\pi}^{\pi} \!\! dq \,
\frac{U_{12} + V_{\mathrm{cos}}(Q,q) + V_{\mathrm{sin}}(Q,q)}
{\mathcal{E}^{(2)} + \epsilon^{(2)}(q) + \epsilon(K-q) - E}
\Psi(q,k), \label{dmPsi}
\ee
where $K = Q+k $ is the total quasi-momentum, 
$V_{\mathrm{cos}}(Q,q) = [2V^{(2)}\cos(q)- 4J \cos(K-q)]\cos(Q)$ and
$V_{\mathrm{sin}}(Q,q) \equiv V_{\mathrm{cos}}(Q,q)$ with $\cos \to \sin$, 
while $U_{12} (\to \infty)$ is an artificial dimer-monomer on-site 
interaction imposing the hard-core condition on Hamiltonian (\ref{effHam}).
Equation~(\ref{dmPsi}) reduces to a non-linear equation for the 
energy $E$ solving which (numerically) we obtain $E_{a1}$ and $E_{a2}$ 
of Fig.~\ref{fig:spectrum}(c). Comparison with Fig.~\ref{fig:spectrum}(a')
reveals good agreement with the exact solution for the three-body problem:
the continuum spectra are practically identical, while there are small
but noticeable differences in the bound-state energies obtained from 
the exact and effective models. These discrepancies are associated
with the internal structure of the dimer \cite{KWEtALPZ,molmer,MVDPdmrs}, 
not accounted for by the effective model. We have verified that with 
increasing the interaction strength $U$, as the two particles forming 
the dimer become tighter co-localized, the discrepancies between
the exact and effective models gradually disappear, as expected 
\cite{MVDPdmrs,DPKLT}.

For $K=0,\pm \pi$, the bound states of Hamiltonian~(\ref{effHam}) 
can be calculated exactly. To this end, we seek the eigenstates of 
$H_{\mathrm{eff}}$ in coordinate basis,
$\ket{\Psi}=\sum_{j_1 \neq j_2} \Psi(j_1,j_2) \ket{j_1,j_2}$,
with the two-particle wavefunction in the form 
$\Psi(j_1,j_2)= e^{iK(j_1+j_2)/2} e^{-i\delta_K j_r} \phi_K (j_r)$, 
where $j_r \equiv j_1 - j_2$, with $j_1$ and $j_2$ being 
the lattice positions of the monomer and dimer, 
and $\tan(\delta_K) = \tan{(K/2)} [J-J^{(2)}]/[J+J^{(2)}]$.   
For the relative coordinate wavefunction $\phi_K (j_r)$, 
imposing the hard-core condition $\phi_K(0)=0$, we then 
obtain the difference relations
\begin{multline}
\bar{J}_K \phi_K(\pm 2) + W_K \phi_K(\mp 1) + 
[\bar{E} - V^{(2)}] \phi_K(\pm 1) = 0 , \\
\bar{J}_K [\phi_K(j_r+1)+\phi_K(j_r-1) ] + 
\bar{E} \phi_K(j_r ) = 0 , \label{phieqs}
\end{multline}
with $|j_r| > 1$, 
$\bar{J}_K \equiv \sqrt{J^2 + J^{(2)2} + 2 J J^{(2)} \cos (K)}$,
$W_K \equiv W \cos (K)$, and $\bar{E} \equiv E - \mathcal{E}^{(2)}$. 
Equation~(\ref{phieqs}) can be solved using the exponential ansatz 
$\phi_K(j_r>0) \propto \alpha_K^{j_r-1}$ and $\phi_K(-j_r) = \pm \phi_K(j_r)$,
which yields $\alpha_K^{(\pm)} = -\bar{J}_K/ [V^{(2)} \mp W_K]$ 
for the symmetric ($+$) and antisymmetric ($-$) wavefunction 
of the bound state ($|\alpha_K| < 1$), with the corresponding energy 
$\bar{E}_{a1(2)} = - \bar{J}_K [1+ (\alpha_K^{(\pm)})^2]/\alpha_K^{(\pm)}$.

It is now easy to see that without the exchange interaction 
there would be no bound states (for any $K$) of the effective 
Hamiltonian (\ref{effHam}). Indeed, this hypothetic ($W =0$) problem
is exactly solvable for all $K$, and for two hard-core bosons 
with nearest-neighbour interaction $V^{(2)}$ there could be only 
one bound state when $|\alpha_K| = |\bar{J}_K/ V^{(2)}| < 1$ \cite{MVDPexH}, 
which is not satisfied in the range of validity of the effective model 
\cite{noBSexch0}. In fact, Hamiltonian (\ref{effHam}) can 
support two bound states because exchange operators preclude 
the Bose-Fermi mapping \cite{girardeau}. The effective nearest-neighbour 
interaction is, however, responsible for the asymmetry in the binding 
energies $E_{B1(2)} = \bar{E}_{a1(2)} \mp 2 \bar{J}_K$ of the exchange-bound 
trimers below and above the continuum $E_{c2}$ [see Fig. \ref{fig:spectrum}(b)].
With increasing the on-site interaction strength $U$, as $V^{(2)} \to 0$, 
we find that the binding energies approach asymptotically the universal limit 
$\lim_{|U|\to \infty} E_{B1(2)} = \mp J/2$. 
In the general case of $|K| \neq 0$, $\pi$, due to the exchange 
interaction, the eigenvalue problem $H_{\mathrm{eff}} \ket{\Psi}= E \ket{\Psi }$ 
does not reduce to a closed expression for the relative coordinate 
wavefunction $\phi_K(j_r)$, and the bound states cannot be calculated 
analytically. However, the universal limits for the binding energies 
$E_{B1(2)} \to \mp J/2$ remain valid for all $K$, since 
$J^{(2)} \to 0$ as $|U/J|\to \infty$. 



\paragraph{Discussion.}

We have found that, in a one-dimensional tight-binding periodic potential,
three strongly interacting bosons can form two families of weakly-bound 
trimers with energies below and above the two-body scattering continuum of 
a single particle (monomer) and an interaction-bound particle pair (dimer).
Intuitively, these trimer states correspond to symmetric and antisymmetric
states of a dimer and monomer at the neighboring lattice sites interacting 
with each other via an effective (particle) exchange interaction. 
In a two-site lattice, the splitting between the two trimer states 
would be equal to twice the exchange rate $W=2J$. The infinite
lattice, however, modifies the energy eigenstates: first a two-body 
continuum, corresponding to the dimer-monomer scattering states,
opens up around the unperturbed energy of the dimer and monomer, 
$[U - 2J^{(2)}] \sim U$, with the width given by the sum of the Bloch 
bands of the two, $4J+4J^{(2)} \gtrsim 4J$ ($J^{(2)} \ll J$ for $U \gg J$). 
This continuum then pushes the modified symmetric and antisymmetric
trimer states away from the edges of the continuum, resulting in their 
binding energies approaching $\mp J/2$ as $U \to \infty$.  

The phenomena discussed above can be observed experimentally 
with cold bosonic atoms in optical lattices \cite{OptLatRev}. 
The preparation of the trimers could be accomplished by loading 
double-wells of a super-lattice with three atoms, followed by 
adiabatic conversion into a homogeneous lattice. 
Using the optical Bragg spectroscopy \cite{BraggSp} 
or lattice modulation rf spectroscopy followed by detection 
of the number of surviving bound pairs \cite{KWEtALPZ}, 
one can then map out the bound trimer and unbound dimer  
energies of Fig.~\ref{fig:spectrum}(a).  

Our effective model involving exchange interaction should be
contrasted with the spin-$\frac{1}{2}$ fermionic case, for which
the Bethe ansatz solution \cite{oneDHub} predicts no three-body 
bound state. Our results, therefore, apply to bosons only \cite{ferTrims}. 
Studying larger number of bosons in a lattice might reveal other 
exotic bound states, while long-range interactions will certainly
play an important role in the formation of bound states 
of three or more bosons in a lattice, as is the case in the two-body 
problem \cite{MVDPexH}.

\begin{acknowledgments}
Useful discussions with Daniel C. Mattis are gratefully acknowledged. 
This work was supported by the EU network EMALI.
\end{acknowledgments}



\begin{thebibliography}{99}
\bibitem{cmntU-U}
The mapping of the spectrum of Hamiltonian (\ref{BHH}) for $U<0$ 
onto that for $U>0$ is achieved by mirror reflection against 
the $E=0$ axis and shift of each particle quasimomentum $k_j$ by $\pi$, 
so that the total quasimomentum is $K \to K + N\pi \; (\mod 2\pi)$.
\bibitem{2bres} 
Two-particle resonances can exist in 1D Hubbard model with nearest 
neighbor interaction \cite{MVDPexH} or in the presence of Feshbach
resonances \cite{fshbchRes}.
\bibitem{noBSexch0}
For $U<0$, at $|K|=0,\pi$ this condition yields, respectively,
$|U|/J < 3/2,11/2$, while the effective Hamiltonian (\ref{effHam})
is valid for $|U|/J > 8$.
\end{thebibliography}
\end{document}